\documentclass[a4paper]{revtex4}
\usepackage{graphicx}
\usepackage{fancyhdr}
\usepackage{amsmath}
\usepackage{color}

\newcommand{\be}{\begin{eqnarray}}
\newcommand{\ee}{\end{eqnarray}}
\newcommand{\nee}{\nonumber\end{eqnarray}}

\newcommand{\drbar}{{\overline{\rm DR}}}

\newcommand{\mch}[1] {m_{\ti \x^+_{#1}}}
\newcommand{\mnt}[1] {m_{\ti \x^0_{#1}}}

\newcommand{\msg}    {m_{\ti g}}
\newcommand{\msu}[1] {m_{\ti u_{#1}}}

\def\gev             {{\rm GeV}}

\def\be            {\begin{equation}}
\def\ee            {\end{equation}}
\def\bea            {\begin{eqnarray}}
\def\eea            {\end{eqnarray}}

\def\a              {\alpha}
\def\b               {\beta}
\def\d               {\delta}

\def\x               {\chi}
\def\ti              {\tilde}
\def\sq              {\ti q}

\def\su                {\ti{u}}
\def\sto                  {\ti{t}}
\def \sca                 {\ti{c}}

\def\dll            {\d^{LL}_{23}}
\def\durr            {\d^{uRR}_{23}}

\def\dulr            {\d^{uLR}_{23}}

\begin{document}

\begin{flushright}
HEPHY-PUB 950/15\\
UWThPh-2015-08\\
\end{flushright}

\title{Impact of quark flavor violation on the decay \boldmath{$h^0(125GeV) \to c\bar{c}$} in the MSSM}


\author{K. Hidaka}
\affiliation{Department of Physics, Tokyo Gakugei University, Koganei, Tokyo 184-8501, JAPAN}
\author{A. Bartl}
\affiliation{Universit\"at Wien, Fakult\"at f\"ur Physik, A-1090 Vienna, AUSTRIA}
\author{H. Eberl}
\affiliation{Institut f\"ur Hochenergiephysik der \"Osterreichischen Akademie der Wissenschaften, A-1050 Vienna, AUSTRIA}
\author{E. Ginina}
\affiliation{Institut f\"ur Hochenergiephysik der \"Osterreichischen Akademie der Wissenschaften, A-1050 Vienna, AUSTRIA}
\author{W. Majerotto}
\affiliation{Institut f\"ur Hochenergiephysik der \"Osterreichischen Akademie der Wissenschaften, A-1050 Vienna, AUSTRIA}
%
%

%\begin{abstract}
%We compute the decay width of $h^0 \to c \bar{c}$ in the MSSM with quark flavor 
%violation (QFV) at full one-loop level adopting the $\overline{\rm DR}$ renormalization scheme.
%We study the effects of $\ti{c}-\ti{t}$ mixing, taking into account the constraints from the B meson data. 
%%
%We show that the full one-loop corrected decay width $\Gamma (h^0 \to c \bar{c})$ 
%is very sensitive to the MSSM QFV parameters.
%In a scenario with large $\ti{c}_{L,R}-\ti{t}_{L,R}$ mixing, $\Gamma (h^0 \to c \bar{c})$ can 
%differ up to $\sim \pm 35\%$ from its SM value.
%%
%After estimating the uncertainties of the width, we conclude that an observation of 
%these SUSY QFV effects is possible at a future $e^+ e^-$ collider such as ILC. 
%\end{abstract}

\begin{abstract}
We compute the decay width of $h^0 \to c \bar{c}$ in the MSSM with quark flavor 
violation (QFV) at full one-loop level in the $\overline{\rm DR}$ renormalization scheme.
We study the effects of $\ti{c}-\ti{t}$ mixing, taking into account the constraints on QFV from the B meson data. 
We find that the full one-loop corrected decay width $\Gamma (h^0 \to c \bar{c})$ 
is very sensitive to the MSSM QFV parameters.
In a scenario with large $\ti{c}_{L,R}-\ti{t}_{L,R}$ mixing, $\Gamma (h^0 \to c \bar{c})$ can 
differ up to $\sim \pm 35\%$ from its SM value.
After estimating the uncertainties of the width, we conclude that an observation of 
these QFV SUSY effects is possible at a future $e^+ e^-$ collider such as ILC. 
\end{abstract}

\maketitle

\thispagestyle{fancy}


%%%%%%%%%%%%%%%%%%%%%%%%%%%%%%%%%%
\section{Introduction}

%It is very important to determine if the SM (Standard Model)-like Higgs boson 
%discovered at the LHC (Large Hadron Collider) in 2012 \cite{Higgs@ATLAS, Higgs@CMS} 
%is the SM Higgs boson or a Higgs boson of New Physics. This is the most important 
%issue in the present particle physics world. 
It is the most important issue in the present particle physics world to determine 
if the SM (Standard Model)-like Higgs boson discovered at the LHC (Large Hadron Collider) 
in 2012 \cite{Higgs@ATLAS, Higgs@CMS} is the SM Higgs boson or a Higgs boson of New Physics. 
In this article based on our paper \cite{Bartl}, 
we study the possibility that it is the lightest Higgs boson $h^0$ of the Minimal Supersymmetric 
Standard Model (MSSM), by focusing on the width of the decay $h^0 \to c \bar{c}$. 
We compute the decay width at full one-loop level in the $\drbar$ renormalization scheme in 
the MSSM with nonminimal Quark Flavor Violation (QFV).

%%%%%%%%%%%%%%%%%%%%%%%%%%%%%%%%%%
\section{Definition of the QFV parameters}

In the super-CKM basis of $\sq_{0 \gamma} =
(\sq_{1 {\rm L}}, \sq_{2 {\rm L}}, \sq_{3 {\rm L}}$,
$\sq_{1 {\rm R}}, \sq_{2 {\rm R}}, \sq_{3 {\rm R}}),~\gamma = 1,...6,$  
with $(q_1, q_2, q_3)=(u, c, t),$ $(d, s, b)$, one can write the squark mass matrices in their most general $3\times3$-block form
\begin{equation}
    {\cal M}^2_{\tilde{q}} = \left( \begin{array}{cc}
        {\cal M}^2_{\tilde{q},LL} & {\cal M}^2_{\tilde{q},LR} \\[2mm]
        {\cal M}^2_{\tilde{q},RL} & {\cal M}^2_{\tilde{q},RR} \end{array} \right),
 \label{EqMassMatrix}
\end{equation}
with $\tilde{q}=\tilde{u},\tilde{d}$. The left-left and right-right blocks in eq.~(\ref{EqMassMatrix}) are given by
\begin{eqnarray}
    & &{\cal M}^2_{\tilde{u},LL} = V_{\rm CKM} M_Q^2 V_{\rm CKM}^{\dag} + D_{\tilde{u},LL}{\bf 1} + \hat{m}^2_u, \nonumber \\
    & &{\cal M}^2_{\tilde{u},RR} = M_U^2 + D_{\tilde{u},RR}{\bf 1} + \hat{m}^2_u, \nonumber \\
    & & {\cal M}^2_{\tilde{d},LL} = M_Q^2 + D_{\tilde{d},LL}{\bf 1} + \hat{m}^2_d,  \nonumber \\
    & & {\cal M}^2_{\tilde{d},RR} = M_D^2 + D_{\tilde{d},RR}{\bf 1} + \hat{m}^2_d,
     \label{EqM2LLRR}
\end{eqnarray}
where $M_{Q,U,D}$ are the hermitian soft SUSY-breaking mass matrices of the squarks and
$\hat{m}_{u,d}$ are the diagonal mass matrices of the up-type and down-type quarks.
%Furthermore, 
%$D_{\tilde{q},LL} = \cos 2\beta m_Z^2 (T_3^q-e_q
%\sin^2\theta_W)$ and $D_{\tilde{q},RR} = e_q \sin^2\theta_W \times$ $ \cos 2\beta m_Z^2$,
%where
%$T_3^q$ and $e_q$ are the isospin and
%electric charge of the quarks (squarks), respectively, and $\theta_W$ is the weak mixing angle.
$D_{\tilde{q},LL}$ and $D_{\tilde{q},RR}$ are the D terms. 
Due to the $SU(2)_{\rm L}$ symmetry the left-left blocks of the up-type and down-type squarks 
in eq.~(\ref{EqM2LLRR}) are related by the CKM matrix $V_{\rm CKM}$.
The left-right and right-left blocks of eq.~(\ref{EqMassMatrix}) are given by
\begin{eqnarray}
 {\cal M}^2_{\tilde{u},RL} = {\cal M}^{2\dag}_{\tilde{u},LR} &=&
\frac{v_2}{\sqrt{2}} T_U - \mu^* \hat{m}_u\cot\beta, \nonumber \\
 {\cal M}^2_{\tilde{d},RL} = {\cal M}^{2\dag}_{\tilde{d},LR} &=&
\frac{v_1}{\sqrt{2}} T_D - \mu^* \hat{m}_d\tan\beta,
\end{eqnarray}
where $T_{U,D}$ are the soft SUSY-breaking trilinear 
coupling matrices of the up-type and down-type squarks entering the Lagrangian 
${\cal L}_{int} \supset -(T_{U\alpha \beta} \ti{u}^\dagger _{R\a}\ti{u}_{L\b}H^0_2 $ 
$+ T_{D\alpha \beta} \ti{d}^\dagger _{R\a}\ti{d}_{L\b}H^0_1)$,
$\mu$ is the higgsino mass parameter, and $\tan\beta$ is the ratio of the vacuum expectation 
values of the neutral Higgs fields $v_2/v_1$, with $v_{1,2}=\sqrt{2} \left\langle H^0_{1,2} \right\rangle$.
The squark mass matrices are diagonalized by the $6\times6$ unitary matrices $U^{\tilde{q}}$,
$\tilde{q}=\tilde{u},\tilde{d}$, such that
\begin{eqnarray}
&&U^{\tilde{q}} {\cal M}^2_{\tilde{q}} (U^{\tilde{q} })^{\dag} = {\rm diag}(m_{\tilde{q}_1}^2,\dots,m_{\tilde{q}_6}^2)\,,
\end{eqnarray}
with $m_{\tilde{q}_1} < \dots < m_{\tilde{q}_6}$.
The physical mass eigenstates $\sq_i, i=1,...,6$ are given by $\sq_i =  U^{\sq}_{i \alpha} \sq_{0\alpha} $.

We define the QFV parameters in the up-type squark sector 
$\delta^{LL}_{\alpha\beta}$, $\delta^{uRR}_{\alpha\beta}$
and $\delta^{uRL}_{\alpha\beta}$ $(\alpha \neq \beta)$ as follows:
\begin{eqnarray}
\delta^{LL}_{\alpha\beta} & \equiv & M^2_{Q \alpha\beta} / \sqrt{M^2_{Q \alpha\alpha} M^2_{Q \beta\beta}}~,
\label{eq:InsLL}\\[3mm]
\delta^{uRR}_{\alpha\beta} &\equiv& M^2_{U \alpha\beta} / \sqrt{M^2_{U \alpha\alpha} M^2_{U \beta\beta}}~,
\label{eq:InsRR}\\[3mm]
\delta^{uRL}_{\alpha\beta} &\equiv& (v_2/\sqrt{2} ) T_{U\alpha \beta} / \sqrt{M^2_{U \alpha\alpha} M^2_{Q \beta\beta}}~,
\label{eq:InsRL}
\end{eqnarray}
where $\alpha,\beta=1,2,3 ~(\alpha \ne \beta)$ denote the quark flavors $u,c,t$.
In this study we consider $\ti{c}_R - \ti{t}_L$, 
$\ti{c}_L - \ti{t}_R$, $\ti{c}_R-\ti{t}_R$, and $\ti{c}_L - \ti{t}_L$ 
mixing which is described by the QFV parameters $\delta^{uRL}_{23}$, 
$\delta^{uLR}_{23} \equiv ( \delta^{uRL}_{32})^*$, $\delta^{uRR}_{23}$, and $\dll$, respectively.
We also consider $\ti{t}_L - \ti{t}_R$ mixing described by the QFC parameter $\delta^{uRL}_{33}$ which is defined 
by eq.~(\ref{eq:InsRL}) with $\alpha= \beta = 3$. 
All QFV parameters and $\delta^{uRL}_{33} $ are assumed to be real.

%%%%%%%%%%%%%%%%%%%%%%%%%%%%%%%%%%
\section{Reference QFV scenario}

\begin{figure}[ht]
\centering
\includegraphics[width=48mm]{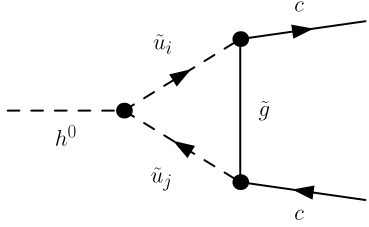}
\caption{Gluino-loop vertex correction to $h^0 \to c \bar{c}$.} \label{gluino_loop}
\end{figure}

%Sample of a table is shown in Table~\ref{example_table}).
%The table caption should appear above the table.
%In the tex file, the labels should be used to refer to tables.
%
We take our reference QFV scenario as shown in Table~\ref{basicparam} \cite{Bartl}. 
The resulting physical masses of the particles are shown in Table~\ref{physmasses}. 
The flavor decomposition of  the lightest up-type squarks $\su_1$ and $\su_2$ is 
shown in Table~\ref{flavordecomp}.
The main features of the scenario are: 
(i) it contains large $\ti c-\ti t$ (scharm-stop) mixings and large QFV trilinear couplings 
of squark-squark-Higgs, and 
(ii) it satisfies the strong constraints on QFV from the B meson data, such as 
BR($b \to s \gamma$), BR($B_s \to \mu^+ \mu^-$) and $\Delta M_{B_s}$, and the 
constraints on the trilinear couplings from the vacuum stability conditions, where scharm 
[stop] is the supersymmetry (SUSY) partner of the charm [top] quark. In this scenario, the lightest 
up-type squarks $\ti u_1$  and $\ti u_2$ are strong mixtures of $\ti c_{L/R}-\ti t_{L/R}$ , 
and the trilinear couplings ($\ti c_L-\ti t_R-h^0$,  $\ti c_R-\ti t_L-h^0$, $\ti t_L-\ti t_R-h^0$ couplings) are large; 
therefore, $\ti u_{1,2}-\ti u_{1,2}-h^0$ couplings are large. This leads to an enhancement of the 
$\ti u_{1,2}-\ti u_{1,2}-\ti g$ loop vertex correction to the decay amplitude of $h^0 \to c \bar{c}$ shown in 
Fig.~\ref{gluino_loop}, where $\ti g$ is a gluino, which is the supersymmetric partner of the gluon. Thus, this results 
in a large deviation of the MSSM prediction for the decay width $\Gamma (h^0 \to c \bar{c})$ from the SM prediction.

%%*****************************************************
\begin{table}[h!]
\caption{Reference QFV scenario: shown are the basic MSSM parameters 
at $Q = 125.5~{\rm GeV} \simeq m_{h^0}$,
except for $m_{A^0}$ which is the pole mass (i.e.\ the physical mass) of $A^0$, 
with $T_{U33} = - 2050$~GeV (corresponding to $\delta^{uRL}_{33} = - 0.2$). All other squark 
parameters not shown here are zero. $M_{1,2,3}$ are the U(1), SU(2), SU(3) gaugino mass parameters.}
\begin{center}
\begin{tabular}{|c|c|c|}
  \hline
 $M_1$ & $M_2$ & $M_3$ \\
 \hline \hline
 250~\gev  &  500~\gev &  1500~\gev \\
  \hline
\end{tabular}
\vskip0.4cm
\begin{tabular}{|c|c|c|}
  \hline
 $\mu$ & $\tan \beta$ & $m_{A^0}$ \\
 \hline \hline
 2000~\gev & 20 &  1500~\gev \\
  \hline
\end{tabular}
\vskip0.4cm
\begin{tabular}{|c|c|c|c|}
  \hline
   & $\alpha = 1$ & $\alpha= 2$ & $\alpha = 3$ \\
  \hline \hline
   $M_{Q \alpha \alpha}^2$ & $(2400)^2~\gev^2$ &  $(2360)^2~\gev^2$  & $(1850)^2~\gev^2$ \\
   \hline
   $M_{U \alpha \alpha}^2$ & $(2380)^2~\gev^2$ & $(1050)^2~\gev^2$ & $(950)^2~\gev^2$ \\
   \hline
   $M_{D \alpha \alpha}^2$ & $(2380)^2~\gev^2$ & $(2340)^2~\gev^2$ &  $(2300)^2~\gev^2$  \\
   \hline
\end{tabular}
\vskip0.4cm
\begin{tabular}{|c|c|c|c|}
  \hline
   $\delta^{LL}_{23}$ & $\delta^{uRR}_{23}$  &  $\delta^{uRL}_{23}$ & $\delta^{uLR}_{23}$\\
  \hline \hline
   0.05 & 0.2 &  0.03   &  0.06  \\
    \hline
\end{tabular}
\end{center}
\label{basicparam}
\end{table}
%
%****************************************************
\begin{table}[h!]
\caption{Physical masses in GeV of the particles for the scenario of Table~\ref{basicparam}.}
\begin{center}
\begin{tabular}{|c|c|c|c|c|c|}
  \hline
  $\mnt{1}$ & $\mnt{2}$ & $\mnt{3}$ & $\mnt{4}$ & $\mch{1}$ & $\mch{2}$ \\
  \hline \hline
  $260$ & $534$ & $2020$ & $2021$ & $534$ & $2022$ \\
  \hline
\end{tabular}
\vskip 0.4cm
\begin{tabular}{|c|c|c|c|c|}
  \hline
  $m_{h^0}$ & $m_{H^0}$ & $m_{A^0}$ & $m_{H^+}$ \\
  \hline \hline
  $126.08$  & $1498$ & $1500$ & $ 1501$ \\
  \hline
\end{tabular}
\vskip 0.4cm
\begin{tabular}{|c|c|c|c|c|c|c|}
  \hline
  $\msg$ & $\msu{1}$ & $\msu{2}$ & $\msu{3}$ & $\msu{4}$ & $\msu{5}$ & $\msu{6}$ \\
  \hline \hline
  $1473$ & $756$ & $965$ & $1800$ & $2298$ & $2301$ & $2332$ \\
  \hline
\end{tabular}
%*********************
\end{center}
\label{physmasses}
\end{table}
%
%****************************************************
\begin{table}[h!]
\caption{Flavor decomposition of $\su_1$ and $\su_2$ for the scenario of Table~\ref{basicparam}. Shown are the squared coefficients. }
\begin{center}
\begin{tabular}{|c|c|c|c|c|c|c|c|}
  \hline
  & $\su_L$ & $\sca_L$ & $\sto_L$ & $\su_R$ & $\sca_R$ & $\sto_R$ \\
  \hline \hline
 $\su_1$  & $0$ & $0.0004$ & $0.012$ & $0$ & $0.519$ & $0.468$ \\
  \hline 
  $\su_2$  & $0$ & $0.0004$ & $0.009$ & $0$ & $0.480$ & $0.509$ \\
  \hline
\end{tabular}
%*********************
\end{center}
\label{flavordecomp}
\end{table}
%}
%****************************************************

%%%%%%%%%%%%%%%%%%%%%%%%%%%%%%%%%%
\section{$h^0 \to c \bar{c}$ at full one-loop level with flavor violation}

We compute the decay width $\Gamma(h^0 \to c \bar{c})$ at full one-loop level in the $\drbar$ 
renormalization scheme taking the renormalization scale as $Q = 125.5~{\rm GeV} \simeq m_{h^0}$ 
in the general MSSM with nonminimal QFV. 
In the general MSSM at one-loop level, in addition to the diagrams that contribute 
within the SM, $\Gamma(h^0 \to c \bar{c})$ also receives contributions from loop 
diagrams with additional Higgs bosons and supersymmetric particles. The flavor violation 
is induced by one-loop diagrams with squarks that have a mixed quark flavor nature. 
The one-loop contributions to $\Gamma(h^0 \to c \bar{c})$ contain three parts, 
QCD ($g$) corrections, SUSY-QCD ($\ti{g}$) corrections and electroweak (EW) corrections 
including loops of neutralinos and charginos.  
In the latter we also include the Higgs contributions. 
Details of the computation of the decay width at a full one-loop level are described in Ref.~\cite{Bartl}. 

%%%%%%%%%%%%%%%%%%%%%%%%%%%%%%%%%%
\section{Numerical results}

% Use the graphics or graphicx packages (distributed with LaTeX2e)
% and the \includegraphics macro defined in those packages.
% See the LaTeX Graphics Companion by Michel Goosens, Sebastian Rahtz,
% entire page. There is no need to do explicit centering.

% \begin{figure}
% \end{figure}

% Surround figure environment with turnpage environment for landscape
% figure
% \begin{figure}
% \end{figure}

\begin{figure}[ht]
\centering
\includegraphics[width=70mm]{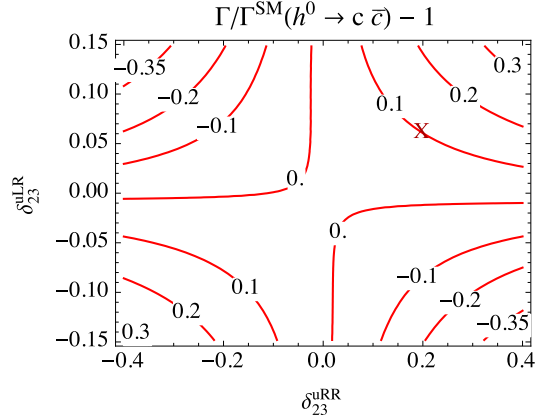}
\caption{
Contour plot of the deviation of the full one-loop level MSSM width $\Gamma (h^0 \to c \bar{c})$ 
from the SM width $\Gamma^{SM} (h^0 \to c \bar{c})$ for our reference QFV scenario of Table~\ref{basicparam}. 
The shown range satisfies all the relevant experimental and theoretical constraints. 
} \label{contour plot of deviation}
\end{figure}

In Fig.~\ref{contour plot of deviation}, we show the contour plot of the deviation of the MSSM prediction from the SM prediction 
$\Gamma^{SM} (h^0 \to c \bar{c})$ = 0.118 MeV \cite{Gamma_SM} in the $\durr$-$\dulr$ plane, where 
$\durr$ and $\dulr$ are the $\ti c_R-\ti t_R$ and $\ti c_L-\ti t_R$ mixing parameters, respectively. 
We see that the MSSM prediction is very sensitive to the QFV parameters $\durr$ and $\dulr$, and 
that the deviation of the MSSM prediction from the SM prediction can be very large (as large as 
$\sim 35\%$). We have found that the MSSM prediction becomes nearly equal to the SM prediction 
if we switch off all the QFV parameters in our reference QFV scenario.

%%%%%%%%%%%%%%%%%%%%%%%%%%%%%%%%%%
\section{Observability of the deviation at ILC}

The observation of any significant deviation of the decay width from its SM prediction indicates 
new physics beyond the SM. It is important to estimate the theoretical and experimental 
uncertainties of the width reliably in order to confirm such a deviation. The relative error 
of the SM width is estimated to be $\sim 6\%$ \cite{Almeida}.  The relative error of the MSSM width 
is also estimated to be $\sim 6\%$ \cite{Bartl}. As seen in Fig.~\ref{contour plot of deviation}, the deviation 
of the MSSM width from the SM width can be as large as $\sim 35\%$. Such a large deviation can be 
observed at a future $e^+ e^-$ collider ILC (International Linear Collider) with a CM energy 500 GeV 
and an integrated luminosity of 1600 $fb^{-1}$, where the expected experimental error of the width is 
$\sim 3\%$ \cite{Tian}. A measurement of the width at LHC is a hard task because of the difficulties 
in charm-tagging.

%%%%%%%%%%%%%%%%%%%%%%%%%%%%%%%%%%
\section{Conclusion}

In this article, we have shown that the full one-loop corrected decay width $\Gamma (h^0 \to c \bar{c})$ 
is very sensitive to the QFV parameters in the MSSM. In a scenario with large $\ti c-\ti t$ mixings, 
the width can differ up to $\sim 35\%$ from its SM value. After estimating the uncertainties of the 
width, we conclude that an observation of these MSSM QFV effects is possible at ILC. Therefore, we 
have a good opportunity to discover the QFV SUSY effect in this decay $h^0 \to c \bar{c}$ at ILC. 

\begin{acknowledgments}
This work is supported by the "Fonds zur F\"orderung der
wissenschaftlichen Forschung (FWF)" of Austria, project No. P26338-N27.
\end{acknowledgments}

\bigskip % extra skip inserted

\end{document}